\documentclass[usenatbib]{mn2e}
\usepackage{aas_macros}
\usepackage{color}

\usepackage[dvips]{graphics}
\usepackage{lscape}

\newif\ifAMStwofonts

\title[SMGs and BzKs]{
A close relationship at $z\sim 2$: submillimetre galaxies and BzK-selected galaxies\footnote{Based in part on data collected at Subaru Telescope through the "Subaru Observatory Project", which is operated by the National Astronomical Observatory of Japan.}
}

\author[T. Takagi et al.]
 {Toshinobu Takagi$^{1}$\thanks{E-mail: takagi@ir.isas.jaxa.jp}, 
Yoshiaki Ono$^2$, Kazuhiro Shimasaku$^{2}$, Hitoshi Hanami$^{3}$ \\
$^1$ 
Institute of Space and Astronautical Science, Japan Aerospace Exploration Agency, 
Sagamihara, Kanagawa 229 8510, Japan\\
$^2$
Department of Astronomy, School of Science, University of Tokyo, 
Tokyo 113-0033, Japan \\
$^3$
Physics Section, Faculty of Humanities and Social Sciences, Iwate
University, Morioka, 020-8550, Japan \\
}

\date{}

\begin{document}

\maketitle

\begin{abstract}
We investigate the relationship between two massive star-forming galaxy populations
at redshift $z\sim 2$; i.e. submillimetre galaxies (SMGs) and BzK-selected galaxies (BzKs). 
Out of 60 SMGs  found in the Subaru/XMM-Newton deep field, 
we collect optical--NIR photometry of 28 radio 
counterparts for 24 SMGs, based on refined sky positions with a radio map 
for 35 SMGs (Ivison et al. 2007). We find a correlation between their $K$-band magnitudes
and $BzK$ [$\equiv (z-K)-(B-z)$] colours: almost all of the $K$-faint ($K_\mathrm{AB} > 21.3$) 
radio-detected SMGs have $BzK>-0.2$, and therefore BzKs. This result gives strong 
support to perform direct optical identification of SMGs by searching for BzKs around SMGs. 
We calculate the formal significance ($P'$ value) for each of the BzK associations 
around radio-undetected SMGs, and find 6 new robust identifications, including one double 
identification. From this analysis, 
we obtain the current best estimate on the surface density of BzK-selected SMGs, which 
indicates that only $\sim1$ per cent of BzKs are SMGs. If BzKs are normal disk-like galaxies 
at $z\sim 2$ as indicated by the correlation between their star formation rate (SFR) and 
stellar mass and also by dynamical properties, SMGs are likely to be merging BzKs. 
In this case, a typical enhancement of SFR due to merging is only a factor of $\sim 3$, which is 
an order of magnitude lower than that of local ULIRGs. This may indicate 
that most of the merging BzKs could be observed as SMGs. Considering a possible 
high fraction of mergers at $z\sim 2$ (at least it would be 
higher than the fraction at $z\la 1$ of $\sim 10$ per cent), 
it is rather puzzling to find such a low fraction of SMGs in the progenitor 
population, i.e. BzKs. 
\end{abstract}


\section{Introduction}\label{sec:introduction}

Among many cosmological surveys, 
the submillimetre (submm) survey is very unique, in the sense that the expected 
flux density of sources is almost insensitive to redshift for $z\approx $ 1 -- 8, 
owing to the strong negative $K$-correction \citep[e.g.][]{2002PhR...369..111B}. 
Although the current sensitivity allows us to detect only the brightest infrared 
galaxies in the universe, it is possible to detect massive starbursts 
and gas rich QSOs at extreme redshifts $z\gg 6$ if exist \citep[e.g.][]{2008MNRAS.383..289P}. 
However, most of the submm galaxies (SMGs) currently identified lie at $z\la 3$. 
This is not because of the detection limit as noted above, but because of 
the `identification limit', owing to a large beam size of current (sub)mm telescopes 
used for surveys. The radio emission provides a high-resolution substitute for 
the infrared emission observed in the submm \citep[e.g.][]{2001ApJ...548L.147C,2003ApJ...585...57C,
2002MNRAS.337....1I,2004MNRAS.355..485B,2004ApJ...606..664D,2007MNRAS.380..199I}, 
and hence the detection limit in the radio has set the upper limits on 
redshifts of identified SMGs \citep{2005ApJ...622..772C}. 
The best way to identify optical counterparts of SMGs is to have high resolution 
submm images with interferometers \citep{2006ApJ...640L...1I,2007ApJ...671.1531Y,2007ApJ...670L..89W,2008ApJ...673L.127D}. Although time consuming, it allows us to perform 
direct optical/near-infrared (NIR) identifications from the submm position. 

Here we adopt another method of direct optical identification using optical/NIR 
colours of galaxies around the submm source. The expected number of random 
optical/NIR associations within the error circle of the submm position is not negligible; 
e.g. there would be $\sim 4$ $K$-band sources ($K_\mathrm{AB} < 23)$ on average within 
a $r<8''$ radius (see below). However, if SMGs have rather confined optical-NIR 
colours and the surface density of objects with similar colours is reasonably low, it 
will be possible to directly identify SMGs without using costly high-resolution submm
images. As in the radio identification method \citep{1986MNRAS.218...31D}, we can calculate 
the formal significance of each optical/NIR association, given the number counts of colour-selected 
galaxies. To do this we need a firm basis for the colours of SMGs. 

Since the discovery of SMGs, extremely red objects (EROs) usually defined with 
$(I-K)_\mathrm{Vega}\ga 4$ or $(R-K)_\mathrm{Vega}\ga 5$ and 
distant red galaxies with $(J-K)_\mathrm{Vega}>2.3$ have been paid special attention 
as candidates of SMG counterparts \citep[e.g.][]{1999MNRAS.308.1061S,2002MNRAS.331..495S,2003ApJ...597..680W,2004MNRAS.354..193C,2004AJ....127..728F,2004MNRAS.351..447C,2005MNRAS.358..149P}. It turns out that the fraction of EROs or DRGs in SMGs is not 
very high and their surface densities are not low enough to reject random associations. 
Among the NIR-selected galaxy populations in the literature, BzK-selected star-forming 
galaxies \citep[BzKs --][]{2004ApJ...617..746D} may be the most promising counterparts of SMGs 
\citep{2005ApJ...633..748R,2007ApJS..172..132B,2007MNRAS.381.1154T}, 
which lie at $1.4\la z \la 2.5$ and 
include heavily obscured galaxies, such as SMGs, as an extreme subset. 

In this paper, we indeed find that almost all of the $K$-faint (indicating higher redshifts) 
radio-detected SMGs are BzKs.
We then use BzKs as a key galaxy population to identify radio-undetected 
SMGs at $1.4\la z \la 2.5$. 
Since the BzK-selection is based on observed wavelengths covering a  
spectral break at 4000\,\AA, we can naturally extend this selection technique to the higher redshift 
range , e.g. $z\gg 3$, in the future by using a combination of wavebands at longer wavelengths.
We also emphasize that the understanding of the physical relation between SMGs and 
BzKs would be important to reveal the evolution of galaxies at $z\sim 2$, given that they 
have similar stellar masses of $\sim 10^{11}$\,M$_\odot$, and spatial correlation lengths 
of $\sim 10$\,$h^{-1}$\,Mpc \citep{2004ApJ...611..725B,2006ApJ...638...72K,2007ApJ...660...72H}.
We discuss the hypothesis of SMGs being merging BzKs.

We adopt the sample of SMGs from the 
SCUBA HAlf Degree Extragalactic Survey \citep[SHADES --][]{2005MNRAS.363..563M,2006MNRAS.372.1621C} as described in Section 2. In Section 3, we investigate typical optical/NIR colours 
of the radio-detected SMGs. We apply our direct identification method for radio-undetected SMGs in 
Section 4. We then discuss a possible evolutionary link between SMGs and BzKs in 
Section 5. Finally we give our summary in Section 6. 
Throughout this paper, we assume a cosmology with
$\Omega_\mathrm{m} =0.3$, $\Omega_\Lambda =0.7$ and 
$H_0 =70$$\,$km$\,$sec$^{-1}$$\,$Mpc$^{-1}$.
All magnitudes in this paper use the AB system unless otherwise noted.

\begin{table*}
\begin{minipage}{120mm}
\tabcolsep=5pt
\caption{Photometry of radio-detected SMGs}
\begin{tabular}{lllcccccccccc}
\hline
  \multicolumn{1}{c}{Name} &
  \multicolumn{1}{c}{R.A.} &
  \multicolumn{1}{c|}{Dec.} &
  \multicolumn{1}{c|}{$K_s$} &
  \multicolumn{1}{c|}{$z'-K_s$} &
  \multicolumn{1}{c|}{$B-z'$} &
  \multicolumn{1}{c|}{$R-K_s$} &
  \multicolumn{1}{c|}{$J-K_s$} \\
  \multicolumn{1}{c}{} &
    \multicolumn{2}{c}{[J2000]} &
      \multicolumn{1}{c}{[AB]$^a$} &
        \multicolumn{4}{c}{[AB]$^b$} \\
\hline
          SXDF850.01&  34.37758&  -4.99355&   22.01$\pm$0.07&    2.59&    1.60&    3.46&    2.54\\
     SXDF850.02&  34.51490&  -4.92432&   21.50$\pm$0.04&    2.61&    1.11&    2.95&    2.13\\
     SXDF850.03&  34.42561&  -4.94124&   18.77$\pm$0.01&    1.02&    2.92&    1.71&    0.55\\
     SXDF850.04&  34.41116&  -5.06093&   21.04$\pm$0.03&    2.83&    1.43&    3.54&    1.37\\
     SXDF850.05&  34.51192&  -5.00856&   20.34$\pm$0.01&    2.26&    2.55&    4.05&    1.50\\
     SXDF850.07&  34.41192&  -5.09108&   21.22$\pm$0.03&    2.67&    1.30&    3.12&    1.75\\
     SXDF850.08&  34.43402&  -4.93019&   22.95$\pm$0.10&    3.59&    0.44&    3.64&  $\ge 0.41^c$\\
     SXDF850.10&  34.60375&  -4.93423&   21.93$\pm$0.05&    2.14&    1.41&    2.69&    1.60\\
     SXDF850.12&  34.49710&  -5.08446&   22.42$\pm$0.05&    1.19&    2.06&    1.64&    0.90\\
     SXDF850.16&  34.55784&  -4.96193&   21.84$\pm$0.03&    2.39&    1.10&    3.17&    1.07\\
     SXDF850.19&  34.61584&  -4.97691&   21.53$\pm$0.04&    2.25&    1.45&    3.17&    0.96\\
     SXDF850.21$^d$ &  34.42711&  -5.07356&   15.51$\pm$0.01&    0.33&    0.85&    0.53&   -0.18\\
     SXDF850.23&  34.42679&  -5.09602&   23.54$\pm$0.12&    2.67&    0.49&    3.15&    1.35\\
    SXDF850.24b&  34.39473&  -5.07499&   20.74$\pm$0.02&    2.24&    2.53&    3.51&    1.16\\
     SXDF850.27&  34.53304&  -5.02927&   21.85$\pm$0.03&    3.01&    1.61&    3.91&    1.33\\
    SXDF850.28a&  34.52867&  -4.98692&   19.20$\pm$0.01&    2.13&    3.26&    3.33&    1.13\\
    SXDF850.28b&  34.52839&  -4.98821&   20.73$\pm$0.01&    2.25&    2.95&    3.99&    1.23\\
     SXDF850.30&  34.41660&  -5.02102&   21.37$\pm$0.03&    3.95&    2.70&    5.29&    2.07\\
     SXDF850.31&  34.39938&  -4.93208&   19.71$\pm$0.01&    1.92&    2.85&    3.00&    1.09\\
     SXDF850.35&  34.50349&  -4.88503&   19.68$\pm$0.01&    1.91&    2.22&    3.04&    1.04\\
     SXDF850.37&  34.35240&  -4.97823&   22.09$\pm$0.07&    3.46&    1.32&    4.35& $\ge 0.73^c$\\
    SXDF850.47a&  34.39322&  -4.98260&   21.00$\pm$0.02&    2.02&    3.52&    4.03&    0.97\\
    SXDF850.47b&  34.39337&  -4.98329&   21.18$\pm$0.02&    1.69&    2.04&    2.89&    0.97\\
    SXDF850.47c&  34.39039&  -4.98269&   19.76$\pm$0.01&    2.13&    1.79&    3.27&    0.93\\
    SXDF850.52a&  34.52133&  -5.08157&   19.12$\pm$0.01&    1.02&    1.70&    1.47&    0.61\\
    SXDF850.52b&  34.52065&  -5.08368&   21.49$\pm$0.02&    2.88&    3.18&    4.83&    1.55\\
     SXDF850.77&  34.40069&  -5.07595&   20.35$\pm$0.01&    1.95&    3.12&    3.36&    1.12\\
     SXDF850.96&  34.50083&  -5.03784&   22.07$\pm$0.04&    1.73&    0.99&    2.41&    1.03\\
\hline\end{tabular}\label{phot_table}
\medskip
Note --- $^a$ Total (Petrosian) magnitudes from the UKIDSS/UDS catalogue. $^b$ Aperture-matched ($2''$) colours, corrected for galactic extinction. $^c$ 5\,$\sigma$ upper limit in $J$. $^d$ A nearby galaxy with large angular size. Colours are derived from total magnitudes. 
\end{minipage}
\end{table*}

\section{Data and Sample} 

\subsection{SMGs with radio identification}
The sample of SMGs is taken from the SHADES source catalogue
 in the Subaru/XMM-Newton Deep Field (SXDF), 
 which contains 60 reliable sources \citep{2006MNRAS.372.1621C}.
\cite{2007MNRAS.380..199I} found robust 41 radio associations for 35 SMGs. 
They calculated the formal significance of each of the potential 
submm/radio associations, by correcting the raw Poisson probability with the method of 
\cite{1986MNRAS.218...31D}. A submm/radio association is regarded as being robust if the corrected 
Poisson probability $P'$ is less than 0.05 with a search radius of 8$''$. 
The detection limit at 1.4\,GHz is 6.7\,$\mu$Jy beam$^{-1}$ in the best region 
(1\,$\sigma$; Ivison et al. 2007) with a FWHM of 1.7$''$. 

One of the SMGs, SXDF850.6, has a triple radio association, 2 robust and 1 tentative. 
A high angular resolution image with Submillimetre Array (SMA) pinned down the 
tentative radio association as a true radio counterpart 
(J. Dunlop et al., private communication). Thus, two `robust' radio associations 
happen by chance. 
Excluding these two chance associations, and adding the true radio counterpart of 
SXDF850.6, we investigate 40 radio associations in total. 
\cite{2007MNRAS.380..199I} also suggest one 
24\,$\mu$m source as a counterpart of SXDF850.71, which has no radio associations. 
We also include this MIR counterpart in the analysis below. However, this source is 
eventually excluded in our study, since we find no reliable $K_s$-band counterpart 
within a search radius of 2$''$. 

\subsection{Optical -- NIR counterparts} 

We search for optical and NIR counterparts of 40 radio and 1 MIR sources with a search 
radius of 2$''$, using the official 
optical source catalogue of SXDS \citep[ver.\,1 --][]{2008arXiv0801.4017F} and 
the NIR source catalogue of UKIDSS data release 2. 
Here we concentrate on the sources detected in the $K_s$ band only. 
The detection limits in the $J$ and $K_s$ bands are 22.61 and 
21.55\,mag \citep[5\,$\sigma$ in Vega;][]{2007MNRAS.375..213W}, respectively. 
In the case of a multiple association for a radio source, 
we adopt the closest $K_s$-band source to the radio position 
as the correct optical-NIR counterpart. 
Some optical counterparts from the SXDS catalogue are found to be different from the 
NIR counterparts (mostly due to confusion of optical sources). 
In this case, we perform aperture photometry 
at the position of NIR sources. 

We use matched aperture photometry with a 2$''$ diameter to determine colours. 
However, when single $K_s$-band magnitudes are quoted they are total magnitudes 
(Petrosian magnitudes in the UKIDSS catalogue). 
We correct optical photometry for the galactic cirrus absorption. 
We note that SXDF850.21 is obviously a nearby galaxy with a large angular size, 
$\sim 30''$. For this galaxy, we measure optical/NIR colours from total magnitudes.  

Using the UKIDSS/UDS catalogue, we find $K_s$-band counterparts for 29 radio 
sources out of 40. For SXDF850.6 (with the SMA identification), we find a $K_s$-band 
counterpart at 0.19$''$ away from the radio position, but we find no optical counterpart, 
although it is very close to an optical source $\sim1.5''$ away. This case is similar to 
three other SMGs discussed in Smail et al. (2004 -- see their Figure 3). 
Excluding this complicated case, we obtain optical--NIR photometry of 28 radio 
counterparts for 24 SHADES sources in the SXDF.  No $K_s$-band source is found at 
the position of the MIR counterpart of SXDF850.71. 
Out of the 28 $K_s$-detected radio sources, 20 are singly associated with the 
SHADES sources (hereafter singly associated SMGs). 
Therefore, these are the most reliable optical counterparts of SMGs. 

The mean separation between radio and 
optical positions is $0.51'' \pm 0.07''$ with 
a maximum of 1.55$''$ (SXDF850.52a, an apparently large galaxy with 2$''$ radius). 
See also \cite{2008arXiv0803.0475C} for discussions on optical identifications 
of the same SHADES sources, in which the goodness of the SED fitting is taken 
into account. Our identifications of corresponding radio sources are consistent 
with those by \cite{2008arXiv0803.0475C}.

We convert $J$ and $K_s$ Vega magnitudes in the UKIDSS catalogue to AB magnitudes with 
$J$\,[AB] = $J$\,[Vega]$+0.889$ and 
$K_s$\,[AB] = $K_s$\,[Vega]$+1.857$\footnote{These conversion factors have been derived from the comparison 
of two catalogues for the same objects in SXDF; one is the UKIDSS/UDS catalogue in the Vega 
system, another is the one used in the analysis similar to this study 
by \cite{2007MNRAS.381.1154T}, originally from the observation with the University of 
Hawaii 2.2\,m telescope with the Simultaneous 3-colour InfraRed Imager for Unbiased 
Survey (SIRIUS; \citealt{2003SPIE.4841..459N}).}.
In order to achieve the same colour selection of galaxies as done in \cite{2004ApJ...617..746D}, 
we correct both $B-z'$ and $z'-K_s$ colours, based on the analysis of 
\cite{2007MNRAS.381.1154T}; i.e. we adopt $(B-z)_\mathrm{corr} = (B-z')+0.2$ and 
$(z-K)_\mathrm{corr} = (z'-K_s)-0.2$. 
Hereafter $(z-K)$ and $(z-K)$ denote the corrected AB colours. 

\subsection{Additional sample from the literature}
In order to give additional support to our results from the 
SHADES/SXDF sample, lacking spectroscopic redshifts so far, we also analyse the SMG sample 
in the Hubble Deep Field \citep[HDF;][]{2005MNRAS.358..149P} with spectroscopic 
redshifts. For this HDF sample, we adopt $B$- and $z'$-band photometry of the 
Subaru/Suprime-cam \citep{2004AJ....127..180C} and $K_s$-band photometry 
from \cite{2005MNRAS.358..149P}. The colours are given by matched aperture 
photometry again, but with a 3$''$ diameter as adopted in \cite{2005MNRAS.358..149P}.
We apply the same colour correction as done 
for the SHADES sample to use the $BzK$ selection criteria, since the 
photometric bands employed are the same. We find that seven SMGs in the HDF have $BzK$ 
colours as well as spectroscopic redshifts. In Section 5, we use a larger sample of SMGs 
from Chapman et al. (2005), including SMGs in the HDF. 
Unfortunately, we can not use this large sample in the main analysis, 
since we find no $z'$-band photometry for the sample as a whole in the literature.

\begin{table}
\begin{minipage}{70mm}
\caption{Surface density of various NIR galaxy populations with $K_s<23$}
\begin{tabular}{lcc}
\hline \hline
  \multicolumn{1}{|c|}{Population} &
  \multicolumn{1}{c|}{Surface density} &
  \multicolumn{1}{c|}{Number per}  \\
  \multicolumn{1}{|c|}{} &
  \multicolumn{1}{c|}{ [arcmin$^{-2}$]} &
  \multicolumn{1}{c|}{8-arcsec radius}  \\
  \hline
$K$-band sources$^a$ &  64.75  &    3.61   \\
BzKs$^b$            & 7.13   &   0.39      \\
EROs$^b$            &3.77   &   0.21   \\
DRGs$^c$             &3.35  &    0.18     \\
$RK$/BzKs$^b$    &1.49  &    0.083    \\
$RJK$/BzKs$^d$          &  0.75    &  0.04       \\
\hline\end{tabular}
\medskip
References --- $^a$\cite{2001PASJ...53...25M}, $^b$Motohara et al. (in prep.),
$^c$ Kajisawa et al. (2006), $^d$ Assumed to be half of $RK$/BzKs.
\end{minipage}
\end{table}

\begin{figure*}
  \resizebox{14cm}{!}{\includegraphics{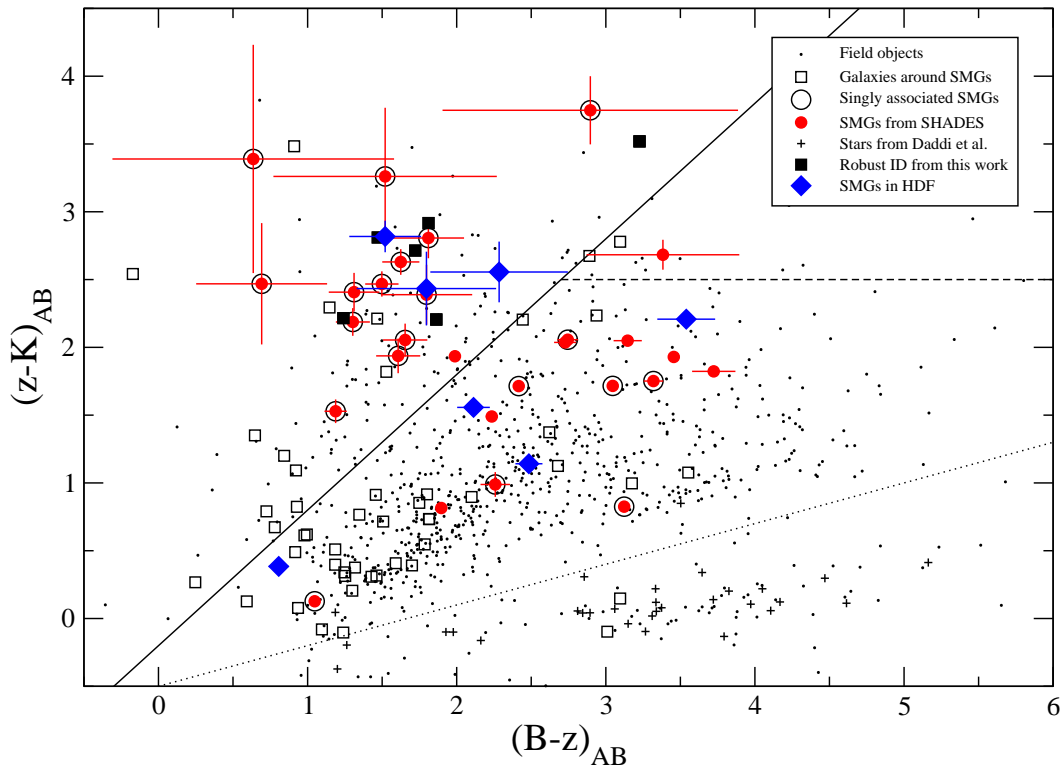}}
 \caption{The BzK colour-colour diagram. Solid circles indicate radio-detected SMGs. 
Encircled ones are singly associated SMGs. Diamonds are the HDF sample from 
Pope et al.\ (2005). Field objects are $K$-band sources in 
the SXDF/SIRIUS field presented in Takagi et al. (2007). Squares indicate $K$-band 
sources within $8''$ radius from radio-undetected SMGs. Filled squares indicate robust 
NIR identification given in Section \ref{sec-id}. Stars from Daddi's catalogue are 
plotted as small pluses, which are used to calibrate the colour correction. 
}
 \label{bzk}
\end{figure*}

\section{Optical-NIR colours of SMG{\sevensize s}}

\subsection{Relation with BzKs}
Figure \ref{bzk} shows the $B-z$ and $z-K$ colours of 28 radio counterparts of SMGs in the 
SXDF including multiple associations, and 7 SMGs in the HDF.
A half (14/28)\footnote{Statistics on SMGs are evaluated for 
the SXDF sample only, because the noise level of the 850-$\mu$m survey 
is uniform in the SXDF, and is not in the HDF.} 
satisfy the colour selection criteria for star-forming 
galaxies at $1.4 < z<2.5$, i.e. $BzK>-0.2$, corresponding to an upper left corner in the 
BzK diagram. Out of the 20 singly associated SMGs, 13 (65$\pm$10\,\%) are classified as BzKs. 

The BzK-selected SMGs have, on average, $(B-z)=1.5$, and $(z-K) = 2.5$. 
Note that the BzK-selected SMGs are confined within colour ranges of 
$1\la (B-z) \la 2$ and $1.5 \la (z-K)\la 3$. Not only SMGs, but also UV-selected (BX/BM) 
galaxies and DRGs have a large overlap with BzKs. While BzKs have 
$\langle z-K \rangle = 1.61$, BX/BM galaxies and DRGs have $\langle z-K \rangle = 1.07$ 
and 2.49, respectively \citep{2005ApJ...633..748R}. 
If BzKs with $(z-K)>3$ are mostly passively evolving galaxies as suggested 
by \cite{2005ApJ...633..748R}, 
SMGs are the reddest star-forming galaxies at $z\sim 2$ as expected from their high 
submm fluxes, indicating a large dust mass and attenuation. 

On the other hand, 
Chapman et al. (2005) found that many of their SMG sample do occupy the BX/BM 
$z\sim 2$ colour, selection region in the $(U-g)$-$(g-R)$ diagram, while SMGs are 
systematically redder in the BzK colours than BX/BM galaxies. This may suggest 
that complex gas/stellar geometry plays a role: when the main body of galaxies is heavily 
obscured, a small fraction of less obscured star-forming regions or 
individual stars can dominate the rest-frame UV fluxes and control the UV colours, while the 
rest-frame optical colours are determined by obscured, but still dominant 
light from the main body. 
Therefore, it is not unreasonable to find galaxies with blue UV and red optical/NIR
colours \citep[e.g.][]{2003PASJ...55..385T}. 

\begin{figure}
  \resizebox{8cm}{!}{\includegraphics{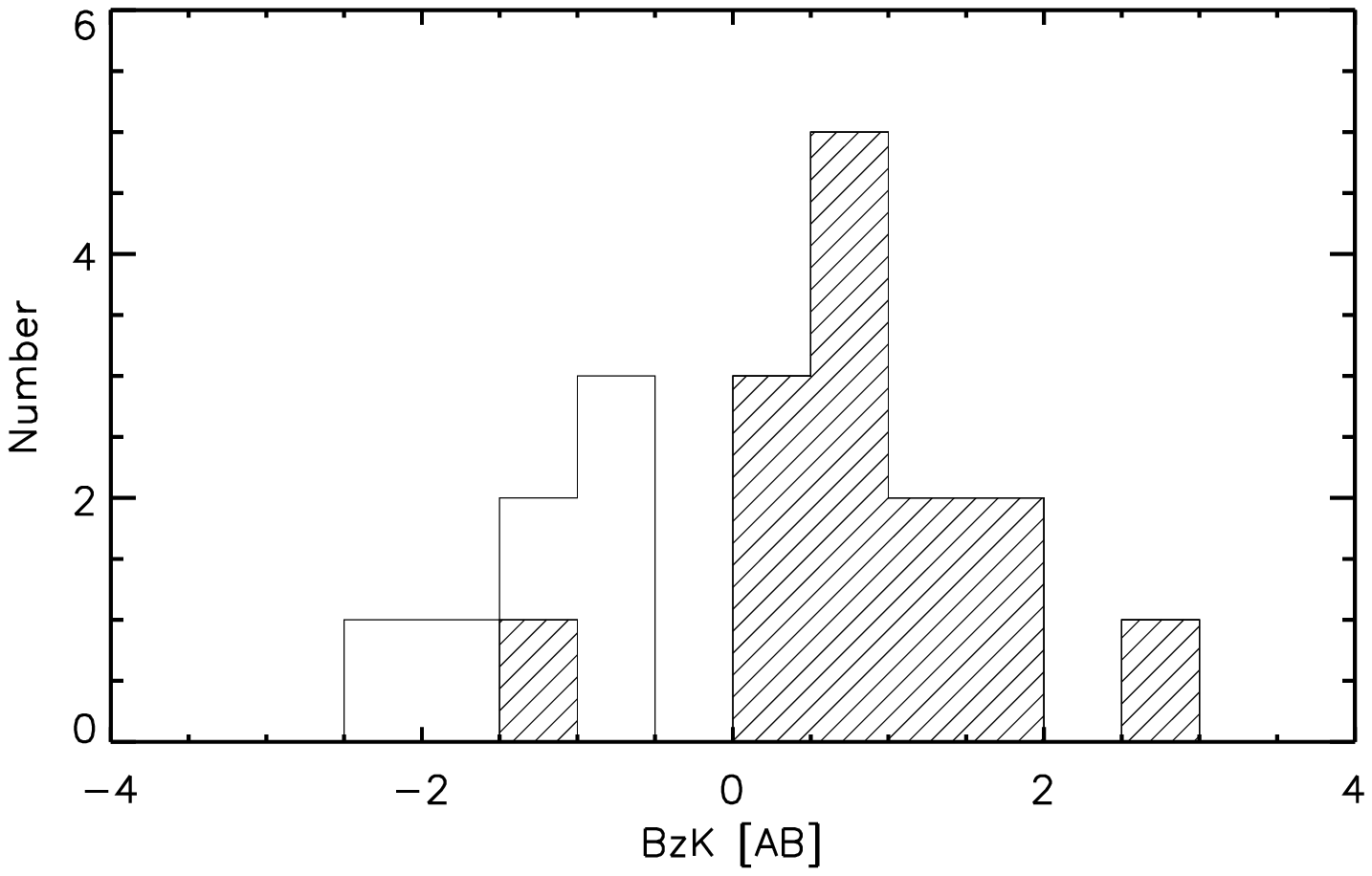}}
  \resizebox{8cm}{!}{\includegraphics{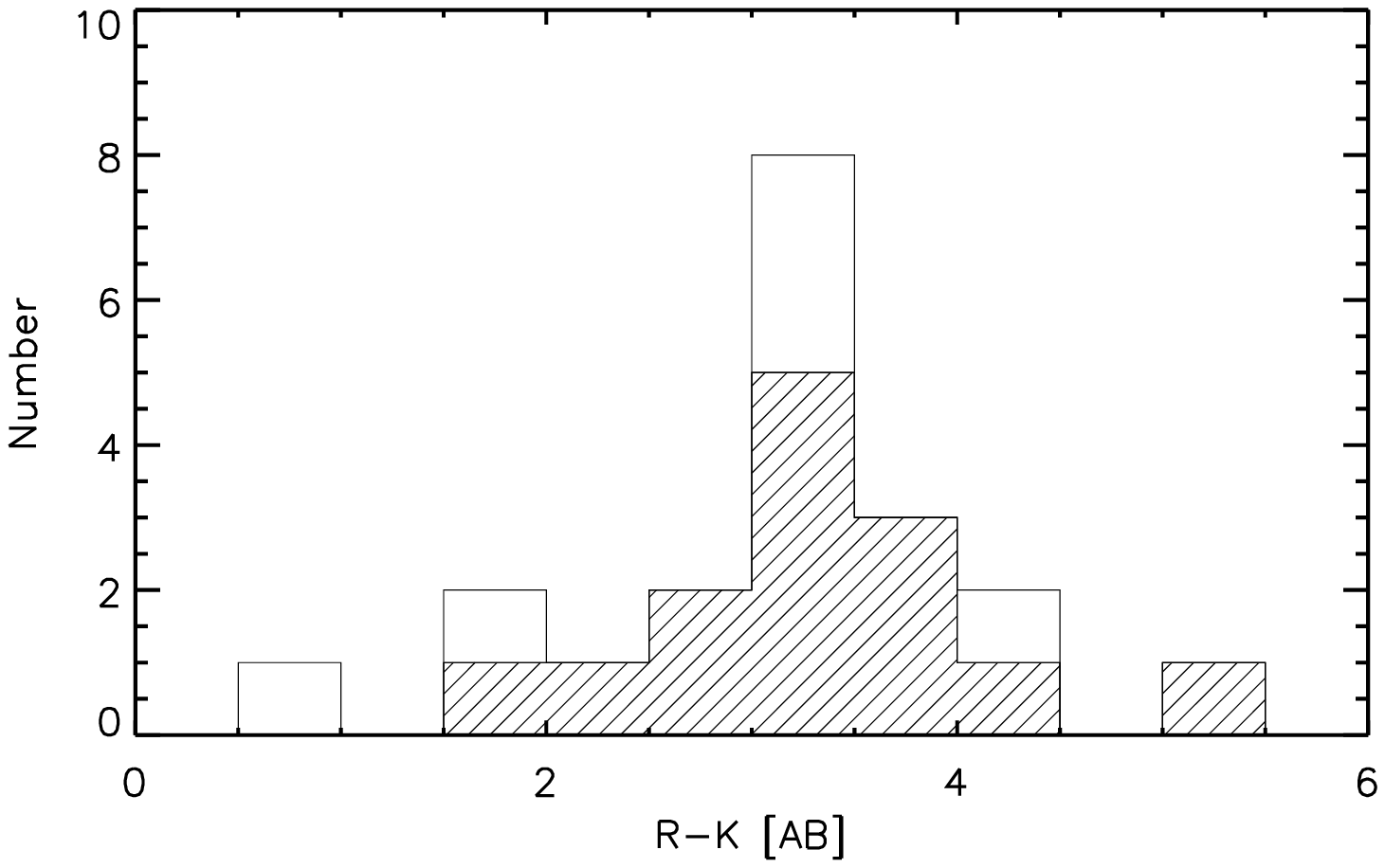}}
    \resizebox{8cm}{!}{\includegraphics{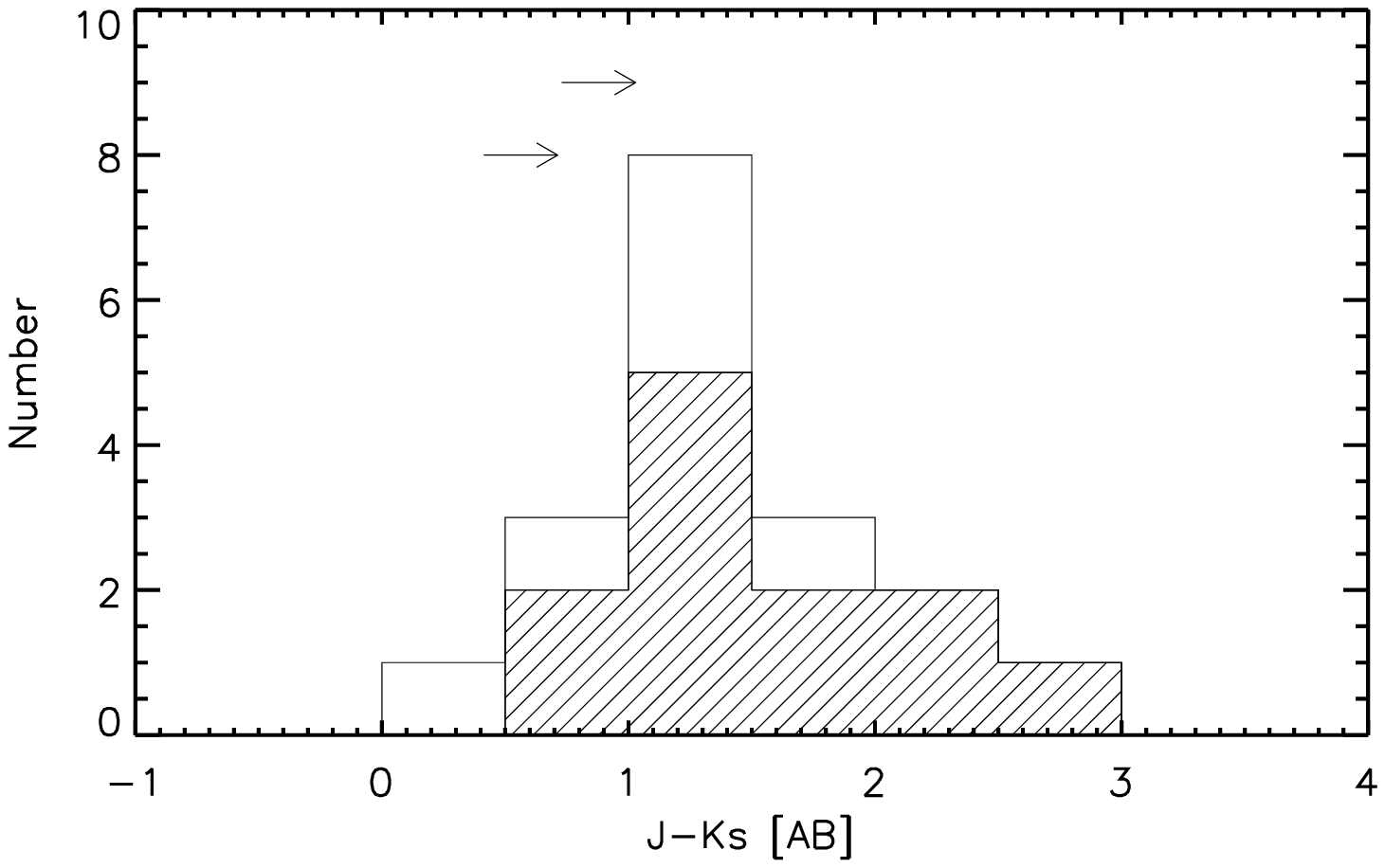}}
 \caption{
Histograms of $BzK$, $R-K$, and $J-K$ colours of singly associated SMGs in the 
SXDF. Hatched histograms indicate $K$-faint ($K_s < 21.3$) galaxies. Distribution of 
the $BzK$ colour is bimodal, and almost all $K$-faint SMGs satisfy $BzK>-0.2$. 
}
 \label{hist}
\end{figure}

As indicated by the error bars in the BzK diagram, most of the BzK-selected SMGs 
belong to an optically fainter sample of SMGs. In order to emphasize this point, 
we show the histogram of $BzK$ (along with $R-K$ and $J-K$) of the singly associated SMGs 
in Figure \ref{hist}.
In these histograms, hatched ones indicate $K$-faint ($K_s > 21.3$) radio-detected SMGs. 
The distribution of $BzK$ is clearly bimodal. This may suggest that BzK-selected SMGs 
have a distinct spectral break at 4000\,\AA, i.e. the Balmer break 
as suggested by \cite{2004ApJ...616...71S}.
We find that 14 singly associated SMGs out of 20 have $K_s>$21.3. Therefore, 
93$\pm$6 per cent (13/14) of the singly associated $K$-faint SMGs are BzKs. 
The HDF sample also follows the same trend.

A high fraction of BzKs in $K$-faint SMGs is implied by the results 
of \cite{2007MNRAS.381.1154T}, using a sub-sample of the SHADES sources. 
Also, \cite{2007ApJS..172..132B} derived a high fraction of BzKs in MAMBO 
1.2\,mm-detected galaxies in the COSMOS field. 
They found that 8/11 (73$\pm$13\,\%) of robust radio counterparts are clearly 
BzKs. One possible difference between 
the MAMBO and SHADES sources could be the fraction of non-BzKs in the samples. 
A half of the radio counterparts of SMGs are non-BzKs, while the MAMBO counterparts 
include only a small fraction of non-BzKs, less than $\sim 20$\% ($<2/11$). 
Since most of the reliable MAMBO counterparts in \cite{2007ApJS..172..132B} 
are faint in the $K$ band ($> 21.3$), 
the difference between MAMBO and SHADES sources could be 
explained by the lack of $K$-bright galaxies in MAMBO counterparts
\cite[see also][]{2004ApJ...606..664D}. This might reflect a systematic difference in the 
redshift distribution between submm (850\,$\mu$m) and millimetre (1.2\,mm) sources. 
This should be tested with more reliable and statistically significant 
(sub)mm surveys with next generation instruments.

\begin{figure}
  \resizebox{8cm}{!}{\includegraphics{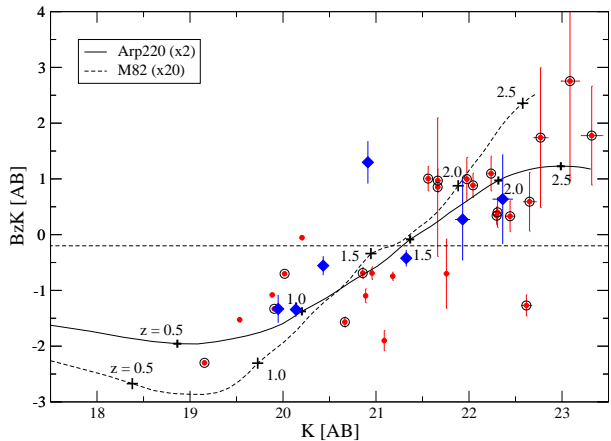}}
 \caption{
$BzK$ colour versus $K$ magnitude for radio-detected SMGs. Symbols are the same as 
in Figure \ref{bzk}. Solid and dashed lines indicate the $BzK$ 
colours and magnitudes of Arp220 and M82 as a function of redshift. 
In order to match the SED templates of these two galaxies 
to the observed trend, their bolometric luminosities 
are multiplied by 2 and 20, resulting in $4\times 10^{12}$ and 
$8 \times 10^{11}$\, L$_\odot$ for Arp220 and M82, respectively. 
}
 \label{colmag}
\end{figure}

\begin{figure}
  \resizebox{8cm}{!}{\includegraphics{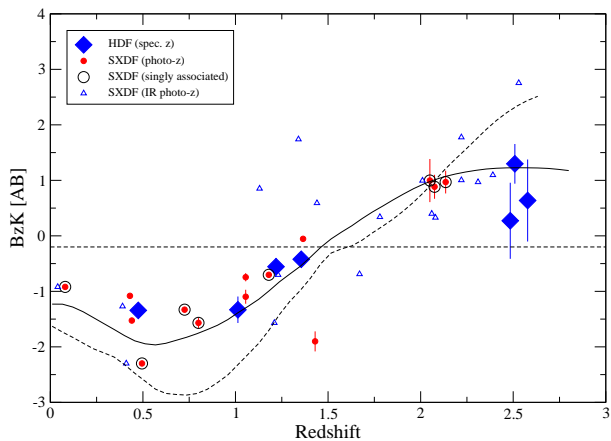}}
 \caption{$BzK$ colour versus redshift for radio-detected SMGs. 
Symbols are the same as in Figure \ref{colmag}. Redshifts of the HDF sample (diamonds) 
are from spectroscopy. For the SXDF sample, we adopt photometric redshifts from 
optical-to-NIR bands (solid circles; Furusawa et al. in preparation) and from optical-to-submm 
bands \citep[triangles;][]{2008arXiv0803.0475C}.
}
 \label{colz}
\end{figure}

\begin{figure}
  \resizebox{8cm}{!}{\includegraphics{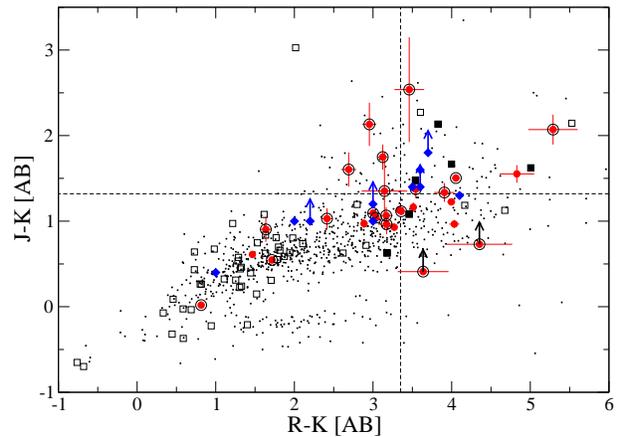}}
 \caption{$J-K$ versus $R-K$ colour-colour diagram. Symbols are the same as in Figure \ref{bzk}.
}
 \label{rjk}
\end{figure}

As expected from the bimodal distribution of the $BzK$ colours in Figure \ref{hist}, there is 
a correlation between $BzK$ colours and $K$-band magnitudes. In Figure \ref{colmag}, 
we show this correlation along with expected colours and magnitudes
of Arp220 and M82, well known nearby dusty galaxies. 
In order to match the SED templates of these two galaxies 
to the observed trend, their bolometric luminosities 
are multiplied by 2 and 20, resulting in $4\times 10^{12}$ and 
$8 \times 10^{11}$\,L$_\odot$ for Arp220 and M82, respectively. 
Both SED templates are consistent with the observed colour-magnitude
relation of SMGs, indicating that the difference in $K_s$-band magnitudes originates
mainly from redshift variations. According to the SED templates, $K$-faint ($K_s > 21.3$), 
BzK-selected SMGs are likely to be at $z\ga 1.5$. 
As originally claimed by  \cite{2004ApJ...617..746D}, the colour 
criterion of $BzK>-0.2$ would be useful to select galaxies having an obscured SED 
such as Arp220 or M82 at $z\ga 1.4$. 

A rough correlation between the apparent $K$-band magnitudes and 
redshifts of SMGs \citep{2003MNRAS.346L..51S,2004ApJ...616...71S} also indicates that $K$-faint 
SMGs have systematically higher redshifts. In Figure \ref{colz}, we show the 
$BzK$ colour versus 
redshift for our SMGs. The redshifts of the HDF sample are spectroscopic ones, while 
we adopt photometric redshifts for the SXDF sample taken from 
Clements et al. (2008 -- from optical-to-submm bands) and 
J. Furusawa et al. (in preparation -- from optical-to-NIR bands). 
The optical-to-NIR photometric redshifts are calculated using a widely-used code 
HYPERZ\footnote{We reject the best-fit solution if the corresponding probability 
is less than 1 per cent.}  \citep{2000A&A...363..476B}. This plot is remarkably similar to 
Figure \ref{colmag}. The $BzK$ colours of SMGs indicate that the Balmer break
of SMGs is strong enough to apply the $BzK$ colour criteria to select SMGs at $1.4\la z \la 2.5$.

In summary, our analysis suggests that $K$-faint SMGs are indeed an extreme subset 
of BzK galaxies at $1.4<z<2.5$, as originally speculated by \cite{2004ApJ...617..746D}.
This result forms the basis of identifying SMGs using 
the $BzK$ colours of associated NIR sources. In order to identify optical counterparts 
through specific galaxy populations (hereafter counterpart populations), 
such as radio sources and BzKs, the surface density of a counterpart population 
should be low enough to ensure a small number of chance associations. 
If we use a counterpart population with a more confined colour space, the number of 
chance associations will be reduced. For this purpose, we focus not only on BzKs, 
but also on EROs and DRGs. Hereafter we refer to the BzK-ERO overlapping population 
as $RK$/BzKs and the BzK-ERO-DRG overlapping population as $RJK$/BzKs.

\subsection{Relation with EROs and DRGs}

Figure \ref{rjk} shows a colour-colour diagram with $R-K$ and $J-K$. 
The colour criteria for EROs and DRGs are indicated by dashed lines. 
The histograms of $R-K$ and $J-K$ of the singly associated SMGs are shown in Figure \ref{hist}. 
Among the 20 singly associated SMGs, 40$\pm$10\,\% (8/20) are classified as EROs with 
$R-K > 3.35$. This fraction becomes 43$\pm 10$\,\% (6/14) for the 
$K$-faint ($K_s > 21.3$) sample. On the other hand, the singly associated SMGs 
include 45 -- 55\,\% (9--11/20) of DRGs having $J-K>1.3$. 
In the $K$-faint sample, there are 
57 -- 71\,\%(8--10/14) of DRGs. The uncertainty in the DRG fraction comes from 
the $J$-band upper limits for two SMGs. There is no significant difference in 
$R-K$ and $J-K$ colours between the $K$-bright and $K$-faint samples. 
Thus, EROs and DRGs themselves are not as particularly useful as BzKs for optical 
identification. 

We identify six $RK$/BzKs, all of which are singly associated SMGs. These galaxies 
have red $J-K_s$ colours satisfying the DRG selection criteria, although two sources have upper 
limits in the $J$ band. Thus, about a half (6/13) of the BzK-selected 
SMGs are found to be EROs and DRGs, i.e. $RJK$/BzKs. 
As discussed in Takagi et al. (2007), this may mean that the duty cycle of 
extremely red SMGs is about a half of the SMG duty cycle. According to 
the radiative transfer SED model for starburst galaxies by \cite{2004MNRAS.355..424T}, 
SMGs could have extremely red colours during the last half of the starburst phase 
because of evolved stellar populations. 

\section{Optical identification of radio-undetected SMG{\sevensize s}}\label{sec-id}

Our analysis shows that almost all of the $K$-faint SMGs satisfy the BzK selection 
criteria $BzK>-0.2$. This gives strong support to perform `direct' optical 
identification of SMGs using BzKs as a counterpart population, when there are no 
$K$-bright associations around SMGs. 
The overlapping populations, such as $RK$/BzKs and $RJK$/BzKs, are useful to 
identify only a limited fraction of SMGs, but could give robust identifications
because of lower surface densities. 
Using BzKs as a counterpart population, 
we employ the same identification procedure as in the radio identification 
\cite[e.g.][]{2002MNRAS.337....1I,2003ApJ...597..680W,2007MNRAS.380..199I}, 
following the method proposed by \cite{1986MNRAS.218...31D}. 
With this method, we could identify radio-faint SMGs at $z\la 2.5$. 
Such SMGs are expected to have a lower dust temperature than radio-detected 
SMGs at a similar redshift range \citep{2005ApJ...622..772C}.

\subsection{Method}
The probability to find a physically unrelated object within a distance $r$ of 
a given submm source may be described by $P_*=1-\exp (-\pi r^2 N_{\ge S_*} )$, where 
$N_{\ge S_*} $ is the 
surface density of galaxies as bright as or brighter than the candidate identification with 
the flux of $S_*$. 
\cite{1986MNRAS.218...31D} describe a correction to this raw Poisson probability, 
considering the contribution from very close associations of fainter sources ($S<S_*$) 
with a probability as low as $P\le P_*$. 
The corrected probability can be written as $P' = 1 - \exp( - E)$ where $E=P_* [1+\ln (P_*/P_c)]$, 
assuming $P \ll 1$. Here $P_c$ is the critical probability to find a source brighter than 
the detection limit within the search radius. 
Because of a rather large search radius of $8''$ in our case, the values of $P_*$ and $P_c$ 
are not small enough to justify the assumption of $P \ll 1$. Therefore, we derive an 
equation valid even for $P \la 1$: $E=y_* [1+\ln (y_*/y_c)]$ where 
$y_{*,c}= -\ln (1-P_{*,c})$. 

Using the same optical-NIR catalogues as used for radio counterparts of SMGs, 
we search for $K$-band sources around the SHADES/SXDF sources with no robust radio 
counterparts with a search radius of $8''$.
We have obtain 56 $K$-band sources around 24 
radio-undetected SHADES sources. These sources are plotted as squares in the 
colour-colour diagrams shown in Figures \ref{bzk} and \ref{rjk}. 
We regard BzKs in this sample as potential optical identifications of radio-undetected  
SMGs. 

We calculate the corrected Poisson probability of chance associations by using the 
number counts of BzKs, $RK$/BzKs (Motohara et al. in preparation) and $RJK$/BzKs. 
Unfortunately, we find no galaxy counts for $RJK$/BzKs in the literature, while 
Takagi et al. (2007) show that 
52$\pm$8\,\% (19/36) of $RK$/BzKs are DRGs at a magnitude limit of $K_s < 22.1$. Also 
\cite{2007MNRAS.379L..25L} report a fraction of 30$\pm$3\,\% (85/283)
using a shallower ($K_s < 21.2$) but wider image. Here we assume that 
a half of $RK$/BzKs are DRGs, i.e. $RJK$/BzKs, at any magnitude bin. This fraction would give
an upper limit on galaxy counts at the bright end. 
These counts are given up to $K_s = 23$.  
For galaxies fainter than this limit, we use the linearly extrapolated 
number counts in $\log N$ versus $K$ magnitude. 
We summarise the surface density of each galaxy population with $K_s < 23$ in Table 2, 
including EROs and DRGs. 

We calculate $P'$ for each of the potential identifications, depending on the colour and 
the $K_s$-band total magnitude of the object under question. 
For each object, we adopt the minimum of $P'$ for the final 
value $P'_\mathrm{NIR}$. 
We regard galaxies with $P'_\mathrm{NIR}<0.1$ as tentative identification, and 
$P'_\mathrm{NIR}<0.05$ as being robust. 
In the presence of associations of $K$-bright sources, we regard any other association
as a tentative identification even with $P' < 0.05$. This is because a half of SMGs are 
expected to be non-BzKs (see Section 3.1), and we have not ruled out the possibility of 
$K$-bright sources being the true counterparts. 

Some SMGs are too faint to be detected in the $K$ band.
In Section 2.2, we find that $\sim 25$\,\% of radio-detected SMGs have no counterparts in 
the $K$-band image. This fraction could be higher for radio-undetected SMGs, since 
they probably include a sub-sample of SMGs lying at the high redshift tail, i.e. $z\ga 3$. 
We caution the readers that even robust BzK identifications are still plausible candidates, 
since we cannot rule out the possibility that true counterparts are not detected in the $K$ band.

\begin{figure}
  \resizebox{8cm}{!}{\includegraphics{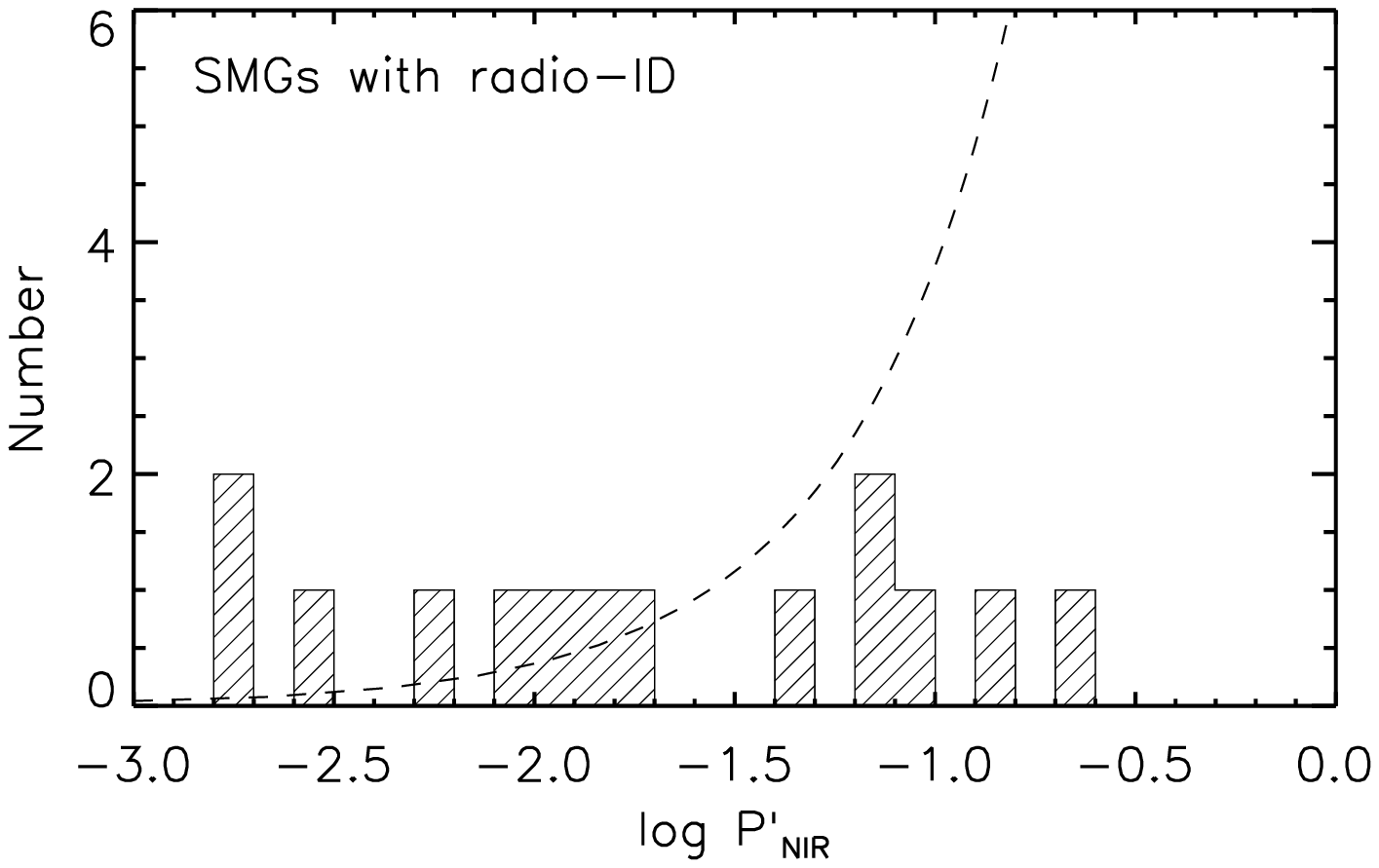}}
    \resizebox{8cm}{!}{\includegraphics{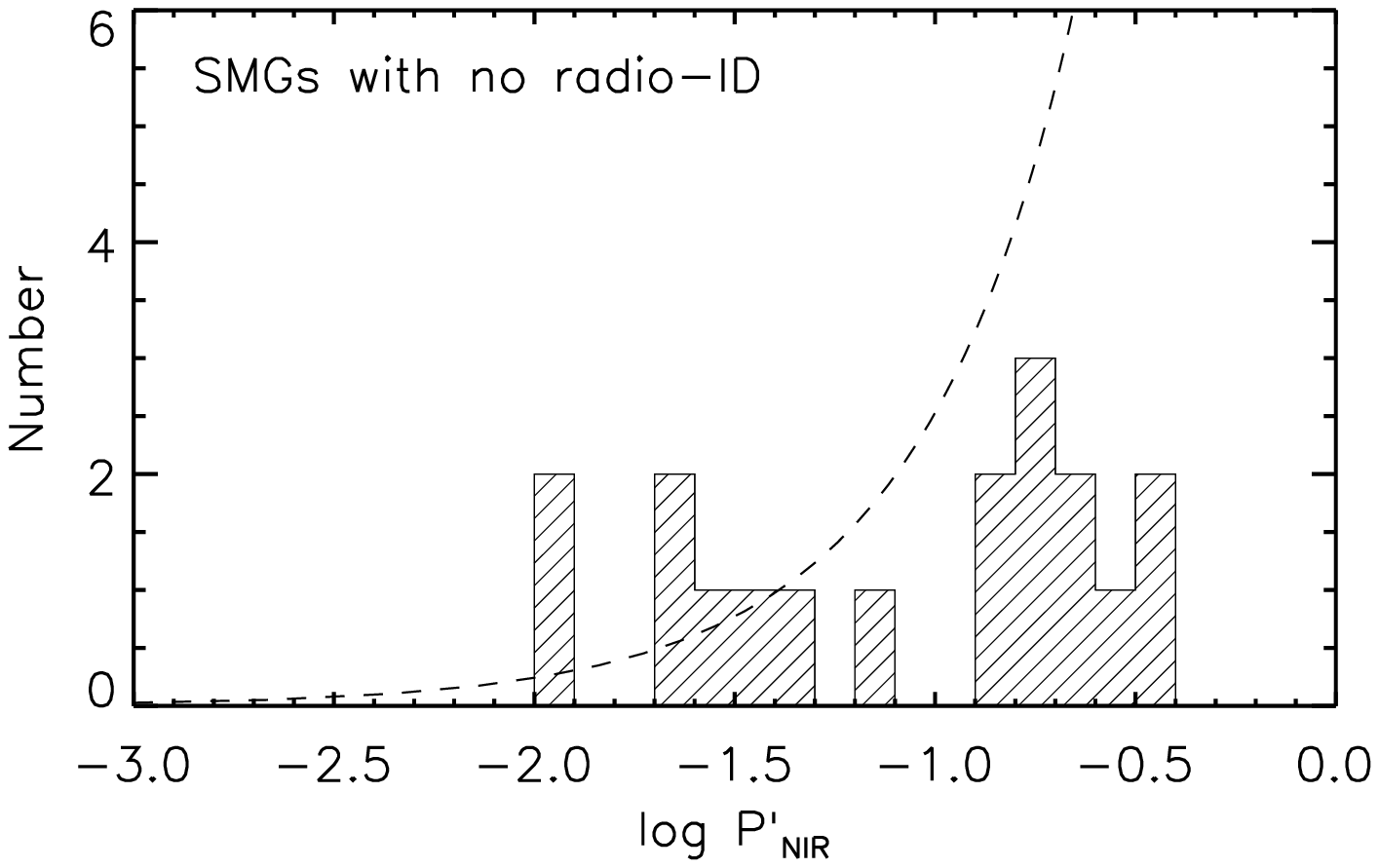}}
 \caption{Histograms of $P'_\mathrm{NIR}$ for radio-detected SMGs and 
for BzKs around radio-undetected SMGs. Dashed line indicate the expected 
number of chance associations as a function of $P'$. See text in detail.
}
 \label{pval}
\end{figure}

\begin{table*}
\begin{minipage}{130mm}
\caption{Optical identification of radio-undetected SMGs}\label{ptab}
\begin{tabular}{lcccccccc}
\hline \hline
  \multicolumn{1}{|c|}{NAME} &
  \multicolumn{1}{c|}{R.A.} &
  \multicolumn{1}{c|}{Dec.} &
\multicolumn{1}{c|}{$K_s$} &
  \multicolumn{1}{c|}{Dist.$^a$} &
  \multicolumn{1}{c|}{$P'^b_\mathrm{BzK}$} &
  \multicolumn{1}{c|}{$P'^c_{RK\mathrm{/BzK}}$} &
    \multicolumn{1}{c|}{$P'^d_{RJK\mathrm{/BzK}}$}  \\
  \multicolumn{1}{|c|}{} &
  \multicolumn{2}{c|}{[J2000]} &
\multicolumn{1}{c|}{[AB]} &
  \multicolumn{1}{c|}{[arcsec]} &
  \multicolumn{1}{c|}{} &
  \multicolumn{1}{c|}{} &
    \multicolumn{1}{c|}{} \\
\hline
   SXDF850.32&  34.34489&  -5.01260&   22.55$\pm$0.06&    7.46&   0.292&   0.076&   {\bf 0.039}\\
   SXDF850.39&  34.46084&  -4.92763&   23.16$\pm$0.12&    0.71&   {\bf 0.020}& ...&  ...\\
   SXDF850.48&  34.35387&  -4.95468&   22.99$\pm$0.09&    4.66&   0.244&   0.057&   {\bf 0.029}\\
   SXDF850.56a$^e$&  34.45994&  -5.10926&   22.48$\pm$0.05&    4.64&   0.182&    0.049&   {\bf 0.025}\\
   SXDF850.56b$^e$&  34.46235&  -5.10814&   21.44$\pm$0.02&    4.96&   0.071&    0.023&   {\bf 0.012}\\
   SXDF850.70&  34.54725&  -5.04682&   21.65$\pm$0.03&    2.54&   0.034&    {\bf 0.011}&  ...\\
   SXDF850.91$^f$&  34.39587&  -4.95518&   23.31$\pm$0.14&    6.08&   0.319&    {\bf 0.075}&  ...\\
   SXDF850.93$^f$ &  34.38840&  -4.96854&   23.76$\pm$0.16&    7.03&   0.316&   0.080&   {\bf 0.041}\\ 
\hline\end{tabular}
\medskip
Note --- Bold figures indicate the adopted $P'_\mathrm{NIR}$. 
$^a$ Angular distance from SHADES centroid. 
$^b$ $P'$ values for BzK galaxies. No entry 
indicates that the galaxy does not satisfy BzK colour criteria. 
$^c$ Same as $^b$ but for the BzK-ERO overlapping population. 
$^d$ Same as $^b$ but 
for the BzK-ERO-DRG overlapping population. 
$^e$ Two robust identifications for one SMG. 
$^f$ Tentative identification, because of $P'>0.05$ (SXDF850.91) or 
the presence of another $K$-bright association (SXDF850.93).
\end{minipage}
\end{table*}

\subsection{Results}

Figure \ref{pval} shows the histogram of $P'_\mathrm{NIR}$ values for BzKs around 
the radio-undetected SMGs in the SXDF and also for the radio identifications. 
The radio identifications have a systematically lower $P'_\mathrm{NIR}$ than 
BzKs around the radio-undetected SMGs, owing to the pre-selection by radio 
with $P'_\mathrm{radio} < 0.05$. 
The dashed lines in Figure \ref{pval} indicate 
the expected number of chance associations as a function of $P'$. The normalization 
of these curves is given by the number of searched regions, i.e. 36 for the SHADES 
sources with radio counterparts and 24 for those without radio counterparts. 

In Table 3, we tabulate the derived $P'$ for BzKs around the radio-undetected SHADES 
sources with $P'_\mathrm{NIR} \le 0.1$. The robust identifications are indicated by solid 
squares in the colour-colour diagrams (Figures \ref{bzk} and \ref{rjk}). 
From the $P'$-statistics of NIR galaxies, we find 6 robust (including one double identification) 
and 2 tentative identifications. Thus, the success rate of identification is not very high, 
only $\sim 15$\,\%. This might be reasonable if we consider the possible bias in radio-undetected 
SMGs toward optically faint counterparts at high redshifts.

All of the robust NIR identifications but one are $RK$/BzKs or $RJK$/BzKs, owing to 
lower surface densities than that of bare BzKs. This indicates that the 
current positional uncertainty is too large to use bare BzKs for a counterpart 
population. Therefore, it is practical to use $RK$/BzKs and/or $RJK$/BzKs, although 
the completeness of identification would decrease by a factor of 2. This is still an 
encouraging performance for future exploitation. We can extend this technique to 
higher redshifts, using a combination of photometric bands at longer wavelengths, 
such as $RJL$ \citep[see][]{2004ApJ...617..746D}, $zKM$ and so on. This technique will 
be useful to push the current `identification limit' of SMGs to $z\gg 3$. 

BzK-selected radio-undetected SMGs would have a lower dust temperature 
than radio detected ones. Although the sample suffers from various selection effects, 
due to optical-NIR colour cuts, we could put some constraints on the number of SMGs with 
cool dust. We identified 5 radio-undetected SMGs as BzKs. On the other hand, 
we found that 13 radio-detected SMGs (singly associated ones) are BzKs. Thus, we 
estimate that the fraction of radio-undetected SMGs at $1.4 \la z \la 2.5$ would be 
$\ga 30$\,\%. This is a lower limit, since we can identify only SMGs with extremely red 
colours such as $RK$/BzKs and $RJK$/BzKs. We also caution that our sample is limited to 
those with $K$-band magnitudes of $K_s \la 23.5$, and missing extremely faint 
SMGs.

\subsection{Fraction of SMGs in NIR galaxy populations}
We summarise the fraction of SMGs in each NIR galaxy population by using the 
observed surface density. The effective area of SHADES in the SXDF 
is $\sim 250$\,arcmin$^2$ \citep{2007MNRAS.381.1154T}. 
Considering both the radio and NIR identifications, we found 
18($+1$), 12($+4$), and 12($+3$) singly associated SMGs (plus possible candidates from multiple identifications) are BzKs, EROs, and DRGs, 
respectively\footnote{Here we count SXDF850.56 having two robust identifications 
as singly associated SMGs, since both are $RJK$/BzK.}. 
These numbers give surface densities of SMGs with BzK, ERO, DRG colours of 
0.07$^{+0.02}_{-0.02}$, $0.05^{+0.03}_{-0.02}$, and 
$0.05^{+0.03}_{-0.02}$\,arcmin$^{-2}$, respectively\footnote{The error includes a contribution 
from multiple identifications and that of survey area.}. 
Using the surface density of each galaxy population 
in Table 2, we derive the fraction of SMGs in BzKs, EROs, and DRGs, to be about 1 per cent. 
These are the current best estimates of the fraction of SMGs in these galaxy populations. 

Nine (6 radio and 3 NIR identified) singly associated SMGs are found to be $RJK$/BzKs, 
giving a surface density of $\sim$0.04\,arcmin$^{-2}$. The surface density of $RJK$/BzKs 
would be around 0.25 -- 0.7\,arcmin$^{-2}$, considering the results by Takagi et al. (2007) 
and Lane et al. (2007). Thus, $\sim$10\,\% of $RJK$/BzKs are likely to be SMGs, and 
therefore $RJK$/BzKs contain a much higher fraction of SMGs than other optical/NIR-selected 
galaxy populations. This confirms the results by Takagi et al. (2007). 

Given the assumed cosmology, a cosmic (co-moving) volume we consider 
is 9$\times 10^5$\,Mpc$^3$ for the redshift range of $1.4<z<2.5$ and 
the survey area of $\sim 250$\,arcmin$^2$. We find 18 BzK-selected SMGs in 
this volume, corresponding to a space density of $2\times 10^{-5}$\,Mpc$^{-3}$.

\section{Discussion} 
We discuss a possible evolutionary link between SMGs and BzKs and its implications 
for physical properties of SMGs, by using the star formation rate (SFR)-stellar 
mass ($M_*$) relation of BzKs. 
In the following, we discuss only SMGs satisfying the BzK selection criterion 
and $K<20$ (Vega), in order to securely derive stellar masses. 
We find seven such SMGs (singly associated ones)
from our sample. In order to increase the sample, we also use the sample of radio-detected SMGs
in \cite{2005ApJ...622..772C}, which have spectroscopic redshifts and $K$-band 
photometry \citep{2004ApJ...616...71S}. We find seven SMGs in their sample, 
which have a starburst-dominated optical spectrum, a redshift at $1.4 \le z \le 2.5$ (i.e. the redshift 
range of BzKs), and $K<20$ (Vega). We refer to this sample and the sample from the SXDF 
as the Chapman sample and the SXDF sample, respectively.

\subsection{Are SMGs merging BzKs?}
The starburst activity of SMGs seems to be induced by galaxy interactions/mergers as indicated 
by their morphology and dynamical properties 
\citep{2004ApJ...611..732C,2006MNRAS.371..465S,2006ApJ...640..228T,
2008arXiv0801.3650T,2007ApJ...671..303B}. Immediate progenitors 
of SMGs would be gaseous star-forming galaxies as massive as SMGs. 
Most plausible candidates for this parent population would be star-forming 
BzKs, given a similar stellar mass and coevality with SMGs. 
Although the SFRs of BzKs are sometimes as high as those of local 
ULIRGs \citep{2005ApJ...631L..13D}, the kinematical properties of BzKs are more similar to  
quiescent disc galaxies
\citep{2006Natur.442..786G,2007ApJ...671..303B}. This quiescent nature of star formation 
in BzKs is also supported by the measurement of the star formation efficiency from the CO 
luminosity, which is an order of magnitude lower than that of SMGs and similar to local 
spirals \citep{2008ApJ...673L..21D}. 
In the following discussion, we assume that some dynamical perturbations occurring in BzKs 
induce the vigorous star formation of SMGs.

\subsection{SFR and stellar mass of SMGs}
We derive the stellar masses of SMGs, using 
the empirical formulae by Daddi et al. (2004, their Eq. 6 and 7 for the SED-fitting--derived mass): 
\begin{eqnarray}
\log (M_* / 10^{11} M_\odot) &= &-0.4(K^{\mathrm{tot}} - K^{11}) + \Delta \log M_*, \\
\Delta \log M_* &=& 0.218 [(z-K)-2.29] 
\end{eqnarray}
where $K^{11}=19.51$ (Vega) is the $K$-band magnitude corresponding on average to a 
mass of $10^{11}$\,M$_\odot$.
Since these formulae are calibrated for $K<20$ (Vega), 
we restrict our sample of SMGs 
to those in this magnitude range as noted above. 
The correction term with $(z-K)$ colour is applied only for the SXDF sample, since 
$z$-band photometry is not available for the Chapman sample in the literature. 
The stellar masses derived from these formulae are based on the Salpeter initial 
mass function extending between 0.1 and 100\,M$_\odot$. 
In deriving the stellar masses, we adopt Petrosian magnitudes from the UKIDSS 
catalogue for estimation of total $K$-band magnitudes. For the Chapman sample, 
we adopt the total $K$-band magnitudes in Smail et al. (2004), which are estimated 
with the aperture photometry of a large (4$''$) diameter. 
We find that the mean stellar mass of 14 SMGs (7 from SXDF and the other 7 from the Chapman 
sample) is 1.2$\times 10^{11}$\,M$_\odot$. If we included SMGs with $K>20$ (Vega) as 
well, the mean stellar mass would reduce about 20\,\%.
The uncertainty in the stellar mass on single objects is about 60\,\% (Daddi et al. 2004). 
However, we caution that the uncertainty may be much larger than this, since SMGs 
would have a large fraction of mass in substantially obscured stellar populations. 
For example, \cite{2005ApJ...635..853B} derive on average 5 times higher stellar 
masses from rest-frame $K$-band magnitudes than those reported from modeling 
the UV/optical SEDs of SMGs in Smail et al. (2004).

The SFRs are derived from radio 1.4\,GHz fluxes and redshifts.
We estimate the FIR luminosities from the radio-infrared luminosity 
relation, $q_L= \log (L_\mathrm{FIR}/ \nu_0 L_\mathrm{1.4\,GHz})$ found for 
SMGs \citep{2006ApJ...650..592K}, 
where $q_L=2.14$ and $\nu_0 = 4.52$\,THz. 
The radio luminosities at the rest-frame 1.4\,GHz are calculated with 
$L_\mathrm{1.4\,GHz} = 4\pi D_L^2 S_\mathrm{1.4} (1+z)^{\alpha-1}$ where 
$D_L$ and $S_{1.4}$ are the luminosity distance and the 1.4\,GHz flux density, 
and $\alpha$ is the spectral index ($S\propto \nu^{-\alpha}$). 
We adopt $\alpha = 0.7$ \citep{1992ARA&A..30..575C}. 
For the SXDF sample, we assume a redshift of $z=1.9$, 
an average redshift of BzKs \citep{2004ApJ...617..746D}.
For the Chapman sample, 
we use the spectroscopic redshifts from Chapman et al. (2005).
We convert the derived FIR luminosity to the SFR using a relation SFR = 
$L_\mathrm{FIR} / 5.8\times 10^9$\,L$_\odot$ \citep{1998ApJ...498..541K}. 
For the combined sample of SMGs, we derive an average SFR
 of 680\,M$_\odot$\,yr$^{-1}$. This becomes 600\,M$_\odot$\,yr$^{-1}$ if we 
 include the SMGs with $K>20$ (Vega).
If we adopt the widely-used local radio-FIR 
 relation by \cite{1992ARA&A..30..575C}, instead of that for SMGs 
 by \cite{2006ApJ...650..592K}, the derived SFRs increases by a factor 
 of $\sim 3$.

Local ULIRGs are also considered for comparison. We adopt the average 
of the sample in \cite{2002ApJ...580...73T}. We use the dynamical mass, 
instead of the stellar mass, which gives a solid upper limit on the stellar mass. 
The averages of the dynamical masses and the SFRs
are 1.3$\times 10^{11}$\,M$_\odot$ and 340\,M$_\odot$\,yr$^{-1}$, 
respectively.

\begin{figure}
  \resizebox{8cm}{!}{\includegraphics{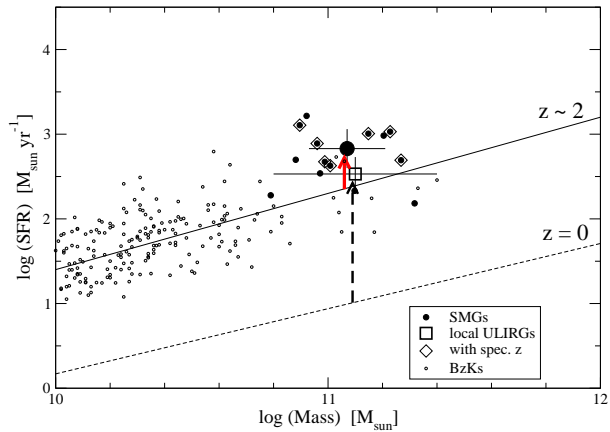}}
 \caption{SFR vs.\ stellar mass of SMGs. For local ULIRGs we adopt the dynamical mass. 
 Solid circles indicate SMGs. 
 Solid circles inside diamonds indicate SMGs with spectroscopic redshifts. 
 Open circles are for BzKs with $K<22$ mag (Vega), 
 taken from Daddi et al. (2007) with MIPS 24$\mu$m-based SFRs.
The large solid circle indicates the average of SMGs, 
whose error bars represent the standard deviation 
 of the combined SMG sample. The large open square corresponds to the average 
 of local ULIRGs. Solid and dashed lines indicate the observed SFR-stellar mass 
 relation of star-forming galaxies at $z\sim 2$ and 0, respectively. 
 Solid and dashed arrows show a typical enhancement of SFR experienced by SMGs and local 
 ULIRGs, respectively. 
}
 \label{mass}
\end{figure}

\subsection{SFR enhancement of SMGs}
We now consider how much SFRs are enhanced in SMGs, owing to dynamical 
perturbations. Such an enhancement may be estimated from a comparison of the average 
SFR of SMGs with that of BzKs, a suspected progenitor population of SMGs, 
with a similar stellar mass. We note that the increase in the stellar mass would be 
less than a factor of $\sim2$ in merging of gas-righ galaxies to produce still gaseous 
SMGs, in which we could neglect the increase in stellar mass after merging. 
The average SFR of progenitor galaxies may be estimated from the correlation between 
SFRs and stellar masses found for BzKs or local star-forming galaxies, once we derive 
the stellar mass of SMGs or local ULIRGs. 

In Figure \ref{mass}, we plot the SFRs and stellar masses of SMGs and local ULIRGs. 
\cite{2007ApJ...670..156D} present the correlation between the SFR and the stellar mass 
of BzKs, SFR = 200$\times M_{11}^{0.9}$ where $M_{11} = 10^{11}$\,M$_\odot$, which is 
depicted as the solid line in the figure. Also, the same correlation but for local blue 
galaxies found in the Sloan Digital Sky Survey -- SFR = 8.7$\times M_{11}^{0.77}$ 
\citep{2004MNRAS.351.1151B,2007A&A...468...33E} -- is shown as the dashed line. 

Comparing the average SFR and $M_*$ of SMGs to the SFR-$M_*$ 
relation of BzKs, we estimate that a typical enhancement of SFR experienced by SMGs 
is around a factor of 3 indicated as the solid arrow in Figure \ref{mass}. 
This enhancement is rather moderate as if SMGs follow the same SFR-$M_*$ relation. 
In the local universe, such a moderate enhancement is, for example, found for M82 
(Figure 18 in Elbaz et al. 2007), whose starburst activity is likely to be induced by interactions 
with M81. Compared to local ULIRGs, this enhancement is less significant by an order of 
magnitude. 
We estimate the SFR enhancement of local ULIRGs to be a 
factor of $\sim 30$ (the dotted arrow), 
comparing the mean SFR and $M_*$ of local ULIRGs with the 
SFR-$M_*$ relation of local star-forming galaxies. This is a lower limit, since 
we use the dynamical mass for local ULIRGs.
If progenitors of SMGs are BzKs with a stellar mass comparable to SMGs,
then the SFR enhancement could {\it not} be as high as local ULIRGs. As \cite{2006ApJ...640..228T}
indicated, SMGs are probably experiencing maximum starbursts \citep{1999ApJ...517..103E}, 
determined roughly by the dynamical time scale of galaxies and negative 
feedback due to super novae. 
The maximum SFR expected for SMGs would be around 600\,M$_\odot$\,yr$^{-1}$ \citep
{2006ApJ...640..228T}, which is already comparable to the observed SFRs. 

What are implications of a moderate ($3\times$) SFR enhancement of SMGs? 
In the local universe, such a moderate enhancement is found for typical interacting pairs of 
galaxies (not necessarily in the most luminous merging phase) 
with a projected distance of $r_p < 20$ kpc \citep{2007arXiv0711.3792L}. 
The numerical simulation by \cite{2007A&A...468...61D} also indicates 
that a typical enhancement of SFR during the major merger phase is only a factor of $<5$. 
If this moderate enhancement is a typical case at $z\sim 2$ as well, 
the fraction of SMGs in star-forming BzKs could be as high as the merger fraction. 
Although the merger fraction at $z>1$ is quite uncertain, it would be at least similar to or 
higher than the value at $z\la 1$ of $\sim 10$\,\% \citep[e.g.][]{2003AJ....126.1183C,2008ApJ...672..177L}, because of 
the hierarchical clustering properties of the galaxy formation process.
This argument leaves us an open question why the fraction of SMGs in BzKs is not
$>10$ per cent, but only $\sim 1$ per cent. 
A bias of the submm surveys, which tend to miss galaxies as luminous as SMGs but 
with higher dust temperatures, should play a role here, 
but the effect would be only a factor of $\sim 2$ \citep{2004ApJ...614..671C}. 
Also we are missing some fraction of radio-undetected, and hence cool-dust SMGs, 
which are not extremely red (see Section 4.2). Even if we consider luminous dusty 
galaxies at $z\sim 2$ with both higher and lower dust temperatures, a correction 
for the incompleteness would be less than an order of magnitude, and therefore not 
enough to reconcile the apparent inconsistency.

We could explain this apparent inconsistency between the rareness of SMGs and its 
moderate SFR enhancement as follows. At $z\sim 2$, galaxy interactions may be
too common, which means that high specific SFRs (mass-normalized SFRs) of BzKs 
are a result of galaxy interactions to some extent. 
In this case, there would be almost no net enhancement of SFR in typical interacting galaxies 
over the SFR-$M_*$ relation. The SFR enhancement of SMGs looks 
moderate, but might be indeed significant as galaxies at $z\sim 2$, possibly caused 
only by rare merging events, such as those with some particular orbit parameters, 
very equal mass ratios, or multiple mergers. For example, according to the numerical simulation by 
\cite{2007A&A...468...61D}, high gas concentrations and hence high SFR 
enhancement can be realized not in direct mergers, but in retrograde mergers.

We should bare in mind that 
SMGs are not the only galaxy population experiencing major mergers. SMGs will 
merely be a tip of iceberg. 
\cite{2007ApJ...656....1L} suggest that UV morphologies of 
BzKs and UV-selected (BX/BM) galaxies are not distinguishable from that of SMGs. 
This indicates either that not only SMGs but also BzKs and BX/BM galaxies are 
experiencing merger-induced starbursts,  or simply that the UV 
morphology is not a reliable indicator of a major merger. 
On the other hand, \cite{2007ApJ...671..303B} show that SMGs are a dynamically 
distinctive galaxy population with large mass concentration, i.e. dynamically hotter 
than BzKs and BX/BM galaxies. Therefore, the UV morphology may not be 
very useful to identify dynamically hot galaxies at $z\ga 2$, as suggested by 
no obvious correlations between the UV morphology and other galactic properties, such 
as SFR, outflow and stellar mass \citep{2007ApJ...656....1L}.
We need more direct measures on dynamical properties of galaxies at $z\sim 2$, in 
order to identify kin of SMGs. 
For this purpose, Atacama Large Millimeter Array \citep[ALMA --][]{2006SPIE.6267E...2B} 
will play an important role, which is capable of measuring gas dynamics of 
$\sim1$-mJy sources, such as BzKs \citep{2005ApJ...631L..13D}.

\section{Summary}
We identify optical-NIR counterparts of 28 radio-detected SMGs, including multiple associations, 
in the SHADES/SXDF, and have compiled optical and near-infrared photometry of these sources. 
We then investigate optical/near-infrared colours ($BzK$, $R-K$ and $J-K$) of radio-detected 
SMGs, along with the additional sample from the HDF, 
to find that almost all of $K$-faint ($K_s >21.3$) SMGs have $BzK > -0.2$, i.e. BzKs. 
This forms a strong basis of identifying radio-undetected SMGs using BzKs around submm sources. 

We calculate the formal significance ($P'$ value) for individual BzKs around radio-undetected 
SMGs, and found 6 robust ($P'\le 0.05$) identifications, including one double identification. 
All of these new identifications but one are 
extremely red, i.e. $RK$/BzKs or $RJK$/BzKs. 
It turns out that the current positional uncertainty is too large to use bare BzKs 
for the direct optical identification. However, this result is encouraging for future exploitation 
of SMGs at $z\gg 3$ using a combination of longer wavebands, such as $RJL$ and $zKM$. 

From a comparison of observed surface densities, 
we find that only $\sim 1$ per cent of BzKs, EROs and DRGs are SMGs, while this fraction 
is as high as $\sim10$\,\% for the BzK-ERO-DRG overlapping population, $RJK/$BzKs. 

The average SFR of 
BzK-selected SMGs is $\sim$680\,M$_\odot$\,yr$^{-1}$. If SMGs are major mergers of 
typical disk-like star-forming galaxies at $z\sim 2$, i.e. BzKs, the enhancement of 
SFR occurring in SMGs would be only a factor of $\sim$3. This enhancement is an order of 
magnitude lower than that of local ULIRGs. Such a moderate enhancement of SFR is found 
in ordinary galaxy pairs with a separation of $<20$\,kpc in the local universe, 
and not significant at all. Thus, 
the rareness of SMGs at $z\sim 2$ is rather puzzling. 
This may be because the SFR of star-forming BzKs is already enhanced by galaxy interactions
to some extent, and hence the moderate SFR enhancement of SMGs over BzKs could 
indeed be significant as a galaxy at $z\sim 2$. 
A quantitative explanation on statistical properties of ULIRGs 
near and far would be a challenging task for the theory of galaxy formation, requiring 
clear understanding of galaxy evolution and the physics of galaxy interactions. 

\section*{Acknowledgments}
We thank all the members of the SHADES consortium for their persistent efforts to publish 
the source catalogue. We are grateful to K. Motohara for providing us the number counts 
of NIR-selected galaxies. Also we are thankful to J. Furusawa for providing us 
photometric redshifts from the SXDS. 
We wish to thank an anonymous referee for very useful and constructive comments, 
including the suggestion to add more samples from the literature.
This work is based in part on data obtained as part of the UKIRT Infrared Deep Sky Survey and 
the Subaru/XMM-Newton deep survey. 
We thank D. Elbaz for providing us compiled spectral energy distributions of Arp220 and M82. 
We also thank E. Daddi for sending us his data points used in Figure 7.
TT acknowledges the Japan Society for 
the Promotion of Science (JSPS -- PD fellow, No.\ 18$\cdot$7747).

\end{document}